\documentclass[twocolumn,aps,pre,floatfix]{revtex4}

\usepackage{graphicx}
\usepackage{pstricks}
\usepackage{pst-node}
\usepackage{amsmath}
\usepackage{amssymb}
\usepackage{epsfig}

\begin{document}

\title{
  Density Analysis of Network Community Divisions
}

\date{\today}

\author{Erik Holmstr\"{o}m}
\email[]{erikh@lanl.gov}
\affiliation{Theoretical Division, Los Alamos National Laboratory, Los Alamos, 
             NM 87545, USA}

\author{Nicolas Bock}
\affiliation{Theoretical Division, Los Alamos National Laboratory, Los Alamos, 
             NM 87545, USA}

\author{Johan Br\"{a}nnlund}
\affiliation{Department of Physics and Astronomy, University of British Columbia, 
             Vancouver, B.C., V6T 1Z1, Canada}

\begin{abstract}

  We present a compact matrix formulation of the modularity, a commonly used
  quality measure for the community division in a network. Using this
  formulation we calculate the density of modularities, a statistical measure of
  the probability of finding a particular modularity for a random but valid
  community division into $C$ communities. We present our results for some
  well--known and some artificial networks, and we conclude that the general
  features of the modularity density are quite similar for the different
  networks. From a simple model of the modularity we conclude that all connected
  networks must show similar shapes of their modularity densities.  The general
  features of this density may give valuable information in the search for good
  optimization schemes of the modularity.

\end{abstract}

\maketitle

\section{Introduction}

The nodes of a network can be grouped into communities which are loosely defined
as groups of nodes that are more ``related'' to each other in some fashion than
they are related to the rest of the network. Such a community division can
reveal important structures of the network. In a recent study, for instance,
\citet{PNAS_101_5241} introduced a method to create a network of gene
co--occurrences from the literature and interpret its communities as groups of
genes related to each other by their function. Since some of the genes in these
communities are not known to be related to the community's function, this method
possibly aids in identifying unknown relationships of this sort.
\citet{PRE_71_046101} used a community analysis on a potential energy landscape
to identify transition states of small Lennard--Jones cluster. Networks have
been very successfully used also to simulate dynamics in various systems.  By
modeling a community structure of individuals using a contact network model,
\citet{JTB_232_71} predicted the dynamics of a SARS outbreak.

Many different approaches have been used to identify community structures in
networks. To name a few more recent methods: vertex similarity
\cite{physics_0510143}, vertex degree gradient \cite{PRE_72_046108}, resistor
network \cite{EPJB_38_331}, Potts Hamiltonian model \cite{PRL_93_218701}, and an
information--theoretic approach \cite{PRE_71_046117}. The most popular methods
appear to be ones based on the network modularity $Q$ introduced by Newman and
coworkers \cite{PRL_89_208701, PRE_67_026126, PRE_69_026113, PNAS_99_7821,
PRE_70_066111}.  The advantage with the modularity $Q$ is that it is a well
defined number that gives the quality of a particular community division in a
network. It is bounded, $-1 \leq Q \leq 1$, and is larger for divisions that
split the network into groups with many intra--edges and few inter--edges
between the groups. 

A number of different strategies have been proposed for finding the optimal
community division based on the modularity. These methods can be broadly
divided into two different classes. Path bound methods are agglomerative or
divisive and either successively add or take away edges in the network so as to
reduce the number of communities by merging existing communities (agglomerative)
or to increase the number of communities by taking away edges and splitting
existing communities (divisive). In both cases, the number of possible community
divisions depend on the previous steps in the algorithm, or the particular path
that was taken in the space of all possible community divisions. The resulting
evolution of the community structure is commonly called a dendrogram. The
different methods in this class differ in the way they identify the edges to be
removed or added.  Examples are the shortest--path betweenness
\cite{PRE_69_026113}, random--path betweenness \cite{PRE_69_026113}, or the
greedy algorithm \cite{PRE_69_066133, PRE_70_066111}.  All these methods have in
common that they follow a dendrogram and attempt to identify the edges to be
removed or added by optimizing the effected modularity change. The number of
communities is changed by at most 1 in each step and only information from the
previous step is used. The quality of these methods is very sensitive to the
strategy employed for identifying the critical edges.

Methods in the second class are not path bound and try to optimize $Q$ directly
without regards to a dendrogram.  Simulated annealing \cite{Physica_A_358_593,
PRE_71_046101, Nature_433_895} is a recent example of one of these techniques,
but other techniques as for instance genetic algorithms are also possible. 

Current results suggest that non path bound optimization strategies outperform
dendrogram bound methods \cite{Duch_05}. However, the number of possible ways of
dividing a network with $N$ nodes into $C$ communities is immense and given by
the Stirling number of the second kind $S_{N}^{C}$ \cite{PRE_69_066133}.  Due to
the discrete nature of how nodes are assigned to communities, the modularity
takes on a discrete set of values. The possibility exists that several divisions
are of similar high quality, or that $Q$ is degenerate as a function of the
community division.  All of these properties of the modularity make it difficult
to optimize the modularity in a non path bound way by using standard
optimization techniques.

In this article we study the properties of the modularity $Q$ in a statistical
sense. Our aim is to gain a deeper understanding of the complexity of network
community divisions in general, and the modularity in particular. By gaining
knowledge of the modularity we believe that it may be possible to find faster
and more accurate community division algorithms. We introduce a matrix algebra
formalism to define a connected community division and obtain the modularity.
We calculate a community division density in the modularity--community space
where we show how the values of $Q$ are distributed in terms of the number of
communities in the division for several different networks.

This article is organized as follows: In section \ref{section:Theory} we will
describe the modularity and introduce a matrix representation of it. In section
\ref{section:Results} we will present and discuss our results and in section
\ref{section:Conclusions} we present our conclusions.

\section{Theory}
\label{section:Theory}

\subsection{Modularity and its Matrix Representation}

A network can be represented by its corresponding \emph{adjacency matrix} $A$.
For a network with $N$ nodes, this matrix is of size $N$ x $N$ where the element
$A_{ij}$ represents the edge between nodes $i$ and $j$. For unweighted networks,
$A_{ij} = 1$ if the edge exists and 0 otherwise. For weighted networks, $A_{ij}
= w_{ij}$, the weight associated with this edge, and in the case of an
undirected network, $A$ is symmetric.  If we do not include self--edges, the
diagonal of $A$ is zero. The underlying network can be divided into $C$
communities, which amounts to labeling each node with one of $C$ community
labels. A compact way of expressing a specific community division is through the
community matrix $P$ which we define as a matrix of size $C$ x $N$, with
elements $P_{ij}$ given by

\begin{equation}
  P_{ij} = 
  \left\{ \begin{array}{lll}
  1 & \mbox{\hspace{0.3cm}} & \mbox{node $j$ is a member of community $i$} \\
  0 & & \mbox{otherwise.}
  \end{array}
  \right.
\end{equation}

\noindent
Newman's assortative mixing matrix \cite{PRL_89_208701}, $e$, can then be
expressed as

\begin{equation}
  e = \left( \sum_{ij} A_{ij} \right)^{-1} P \,\, A \,\, P^{T},
  \label{eq:assortative_mixing_matrix}
\end{equation}

\noindent
where $P^{T}$ is the transpose of $P$. The modularity is given by
\cite{PRE_67_026126, PRE_69_026113}

\begin{equation}
  Q = \mathrm{Tr} (e) - \sum_{ij} (e^2)_{ij}.
  \label{eq:modularity}
\end{equation}

\noindent
The larger the value of $Q$ the better the community division. The modularity
has the property that it has an upper bound, $Q \leq 1 - 1/C$. This has to be
regarded as a theoretical upper bound, however; in practice the upper bound is
lower.

\subsection{Statistical Analysis of the Modularity}

\begin{figure}
  \epsfig{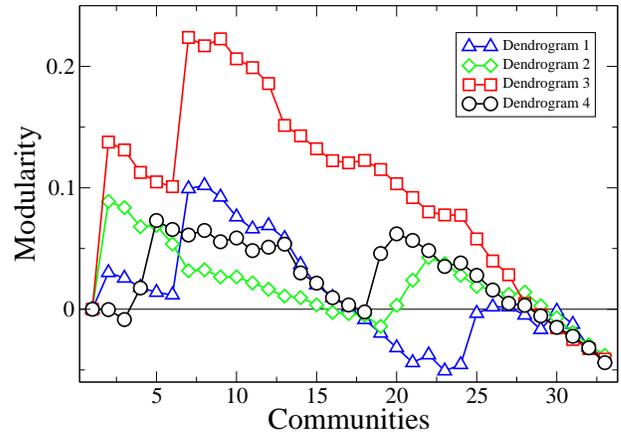}
  \caption{
    (Color online) A set of modularity plots obtained from random dendrogram
    walks in the Zachary Karate Club friendship network. On the x--axis is the
    number of communities, $C$.
  }
  \label{fig:Random-Shot}
\end{figure}

The modularity can be interpreted as a function of the community matrix $P$. In
the space of all possible community divisions, $Q (P)$ defines a rugged and
complicated surface.  In Fig.\ \ref{fig:Random-Shot} we show several curves of
$Q(P)$ vs. $C$ obtained by randomly choosing a path through this space along a
dendrogram in the Zachary Karate Club network \cite{J_Anthropol_Res_33_452}. The
path is chosen by starting with a diagonal $N$ x $N$ matrix $P$ so that all
nodes are in their own community. Then, by summing two randomly selected rows in
$P$ we merge two of the communities.  We can easily check that the new community
is connected by checking the assortative mixing matrix $e$.  We can see in the
figure that the qualities of the various community divisions depend strongly on
the chosen paths and the success in the previous steps.


In the following we will try to gain a more general understanding of the
structure of $Q(P)$ for different networks. We will not attempt to optimize the
modularity but rather map its structure in the space of possible community
divisions.
Put in other words, we would like to know how many community matrices $P$ exist
for a given number of communities $C$ in a given modularity interval between $Q$
and $Q + \delta Q$.
If we start out with the completely split network, i.e. $C = N$, and sample
$Q(P)$ along a random dendrogram until we reach $C = 1$ we will find that some
values of $Q_{C}$ are more likely to be found this way than others.  After
sampling a large number of these random dendrograms we can analyze the result as
a frequency distribution $f(Q)$ vs. $Q$ and $C$ and get a density of
modularities, $N (Q)$. We expect the following: For a given number of
communities, $C$, there will be a range of possible values of $Q_{C}$. We know
from previous studies that it is difficult to find a division with a large
modularity. This implies that it is unlikely for us to find a large value of $Q$
with our random dendrograms and likely that we find some average $Q$.

\section{Results}
\label{section:Results}

\subsection{Examples of modularity densities}

\subsubsection{Real networks}

In Fig.\ \ref{fig:Random-Shot} we present plots of the modularity as a function
of $C$ along random dendrograms in the Zachary Karate Club network
\cite{J_Anthropol_Res_33_452}. The modularities in these examples are quite
different. By performing a large set of such dendrogram walks where we save each
modularity plot we obtain a statistical image of the modularity. 

\begin{figure}
  \epsfig{file=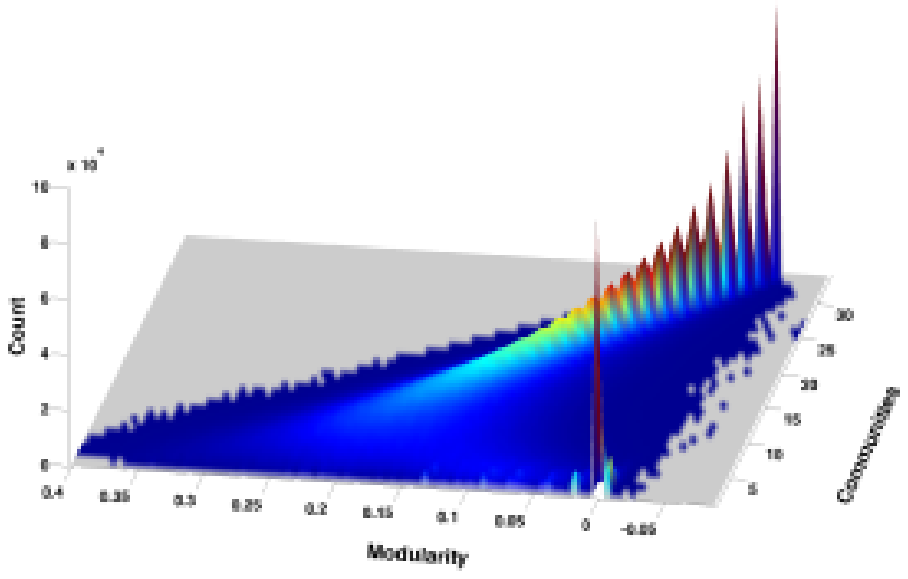,width=0.45\textwidth}
  \epsfig{file=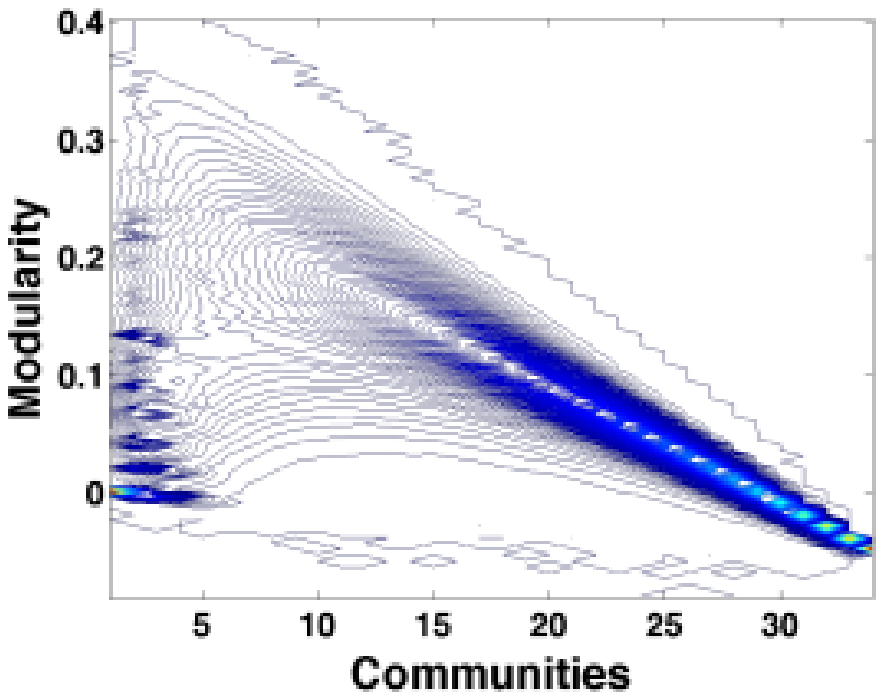,width=0.45\textwidth}
  \caption{
    (Color online) Surface and contour plots showing the distribution of random
    dendrograms for the Zachary Karate Club friendship network. The number of
    dendrograms in this figure is 100,000 and the modularity bin size is 0.01.
    On the x--axis is the number of communities, $C$.
  }
  \label{fig:Zachary-Surf}
\end{figure}

In Fig.\ \ref{fig:Zachary-Surf} we show the modularity density for the Zachary
Karate Club network where we have calculated 100,000 modularity plots from
random dendrogram walks. We find that the modularity density is not uniformly
distributed in the modularity--communities space and has a strong peak for large
values of $C$. This peak decreases rapidly as the number of communities is
decreased. For small values of $C$ the density is low and spread over a large
range of modularities. By construction, the integral over modularity for
constant number of communities is always the same.  Consequently, if the peak is
very low, the range over which the modularity density is spread out will be
large and it thus seems most probable to find the maximum $Q$ within this
regime. 

We notice that although the density is very low in the case of a small number of
communities the peak does not disappear and the top of the peak outlines a
curved shape in the modularity--communities space. The position of this peak
determines the most probable $Q(C)$ relation for the network.

\begin{figure}
  \epsfig{file=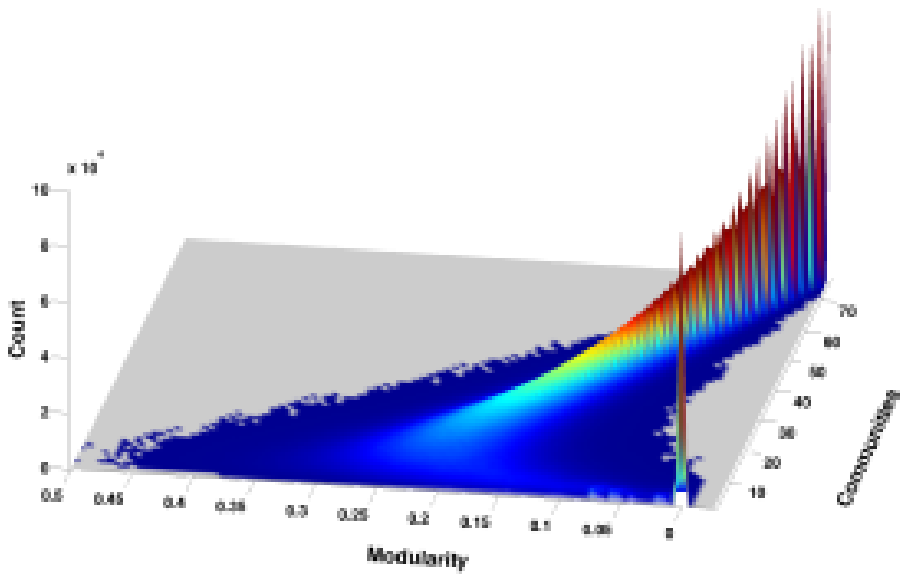,width=0.45\textwidth}
  \epsfig{file=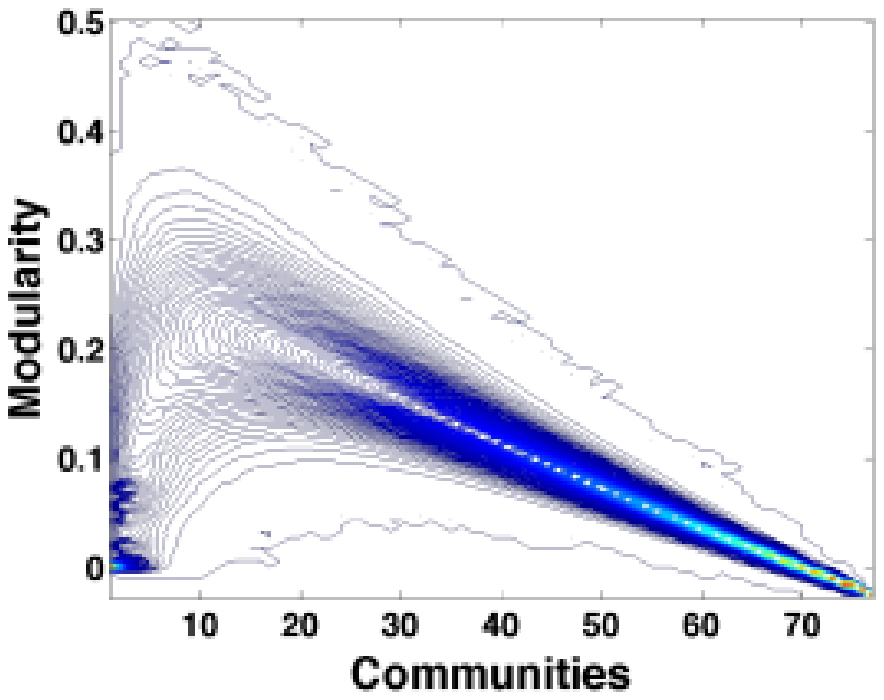,width=0.45\textwidth}
  \caption{
    (Color online) Surface and contour plots showing the distribution of random
    dendrograms for the Les Mis\'{e}rables network. The number of dendrograms in
    this figure is 100,000 and the modularity bin size is 0.01.  On the x--axis
    is the number of communities, $C$.
  }
  \label{fig:LesMiserables-Surf}
\end{figure}

In Fig.\ \ref{fig:LesMiserables-Surf} we show the modularity density surface for
the network of simultaneous appearance on stage for the characters in the Les
Mis\'{e}rables musical \cite{LesMis, PRE_69_026113}. We find that the shape and
general features are similar to the Zachary modularity density surface. The main
differences are that the Les Mis\'{e}rables surface shows smaller regions of
negative modularity and a more shallow curvature. In these two examples, we find
that the general features of the modularity density are very similar. Therefore
we will investigate the modularity density of a few artificial networks to see
whether the same general features can be found.

\begin{figure}
  \epsfig{file=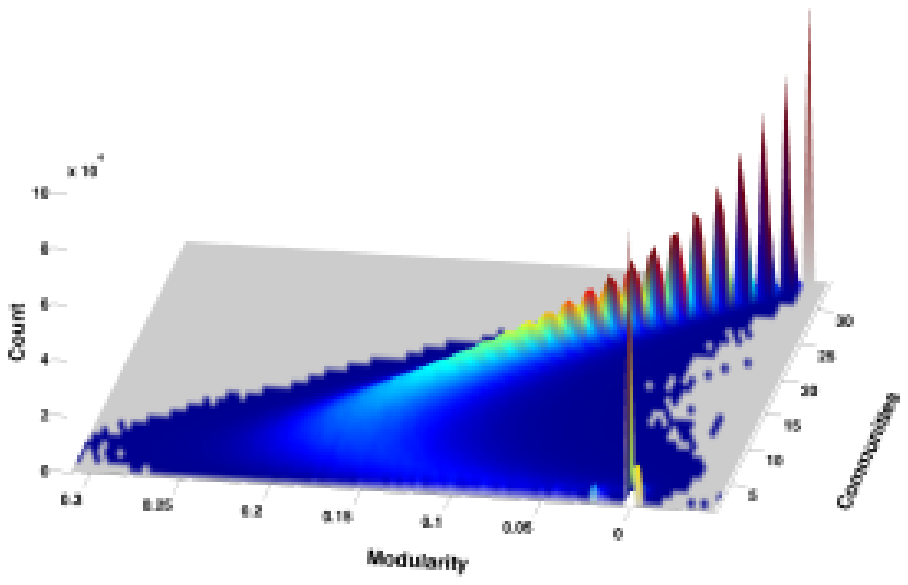,width=0.45\textwidth}
  \epsfig{file=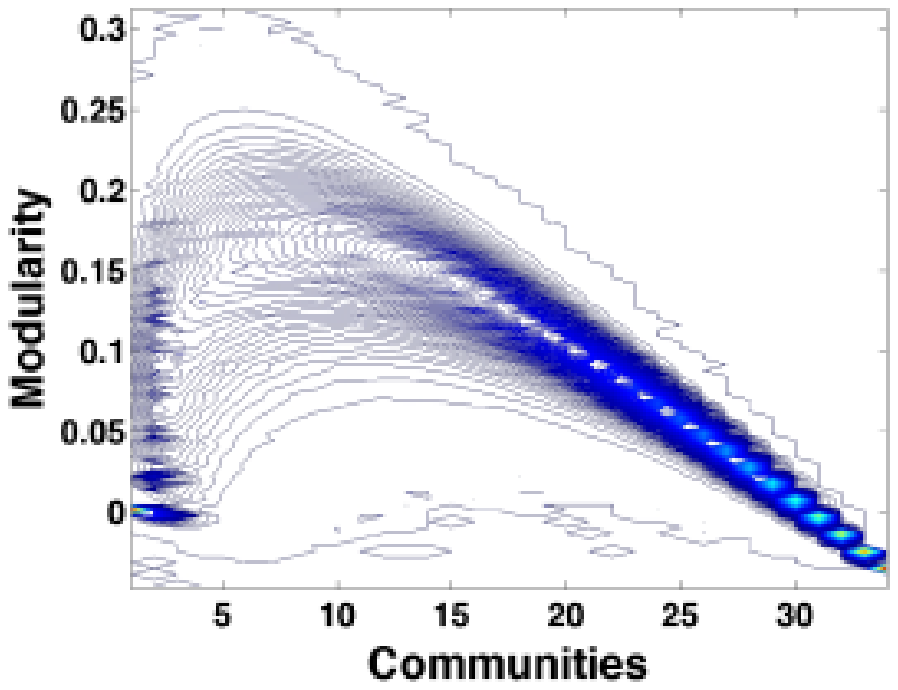,width=0.45\textwidth}
  \caption{
    (Color online) Surface and contour plots of random dendrograms for a random
    network with 34 nodes and the same number of edges (but randomly
    distributed) as the Zachary Karate Club.  On the x--axis is the number of
    communities, $C$.
  }
  \label{fig:random_1}
\end{figure}

\subsubsection{Random networks}
\label{sec:random_networks}

In Fig.\ \ref{fig:random_1} we show the modularity density for a random network
with 34 nodes and 78 undirected edges, which is the same average degree ($\left<
k \right> = 4.59$) as the Zachary Karate Club. The edges are randomly
distributed however and the network does not exhibit any particular community
structure and certainly not the same community structure as the Zachary network.
The modularity density on the other hand does exhibit a very similar structure
to what we found for the networks of Figs.\ \ref{fig:Zachary-Surf} and
\ref{fig:LesMiserables-Surf}. This indicates that to a large extent the
structure of the modularity density is not associated with the particular
community structure, but rather with the network itself.

\begin{figure}
  \epsfig{file=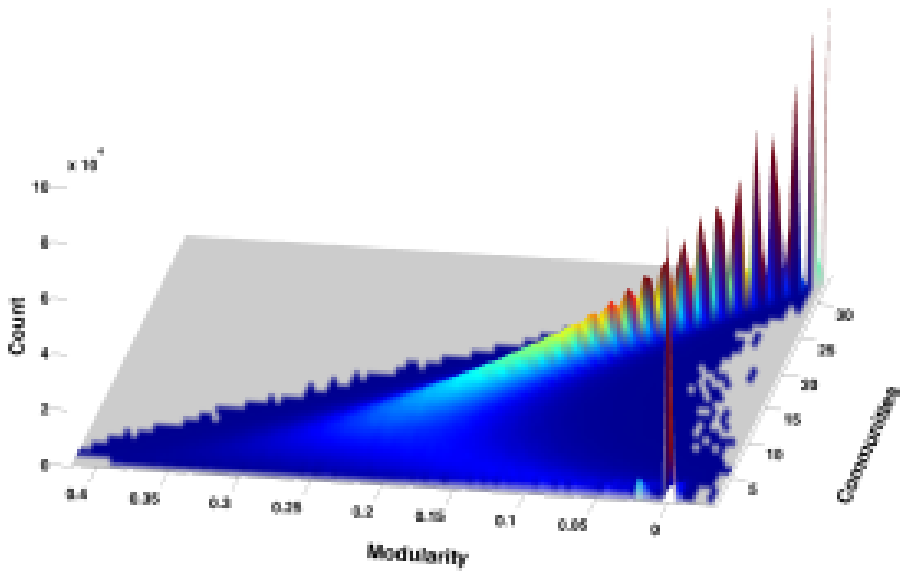,width=0.45\textwidth}
  \epsfig{file=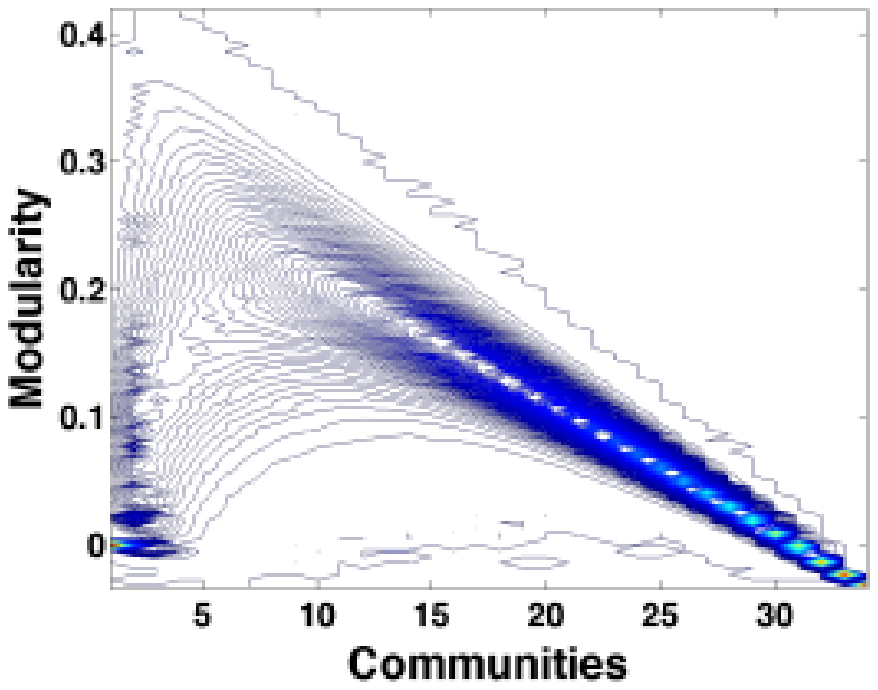,width=0.45\textwidth}
  \caption{
    (Color online) Modularity density for a random network with 34 nodes and 78
    edges. The nodes were randomly assigned to 2 communities and the edges were
    randomly distributed under the constraint that the probability of connecting
    two nodes within the same community is 10 times larger than the probability
    of connecting two nodes which are in different communities.  On the x--axis
    is the number of communities, $C$.
  }
  \label{fig:random_2}
\end{figure}

\begin{figure}
  \epsfig{file=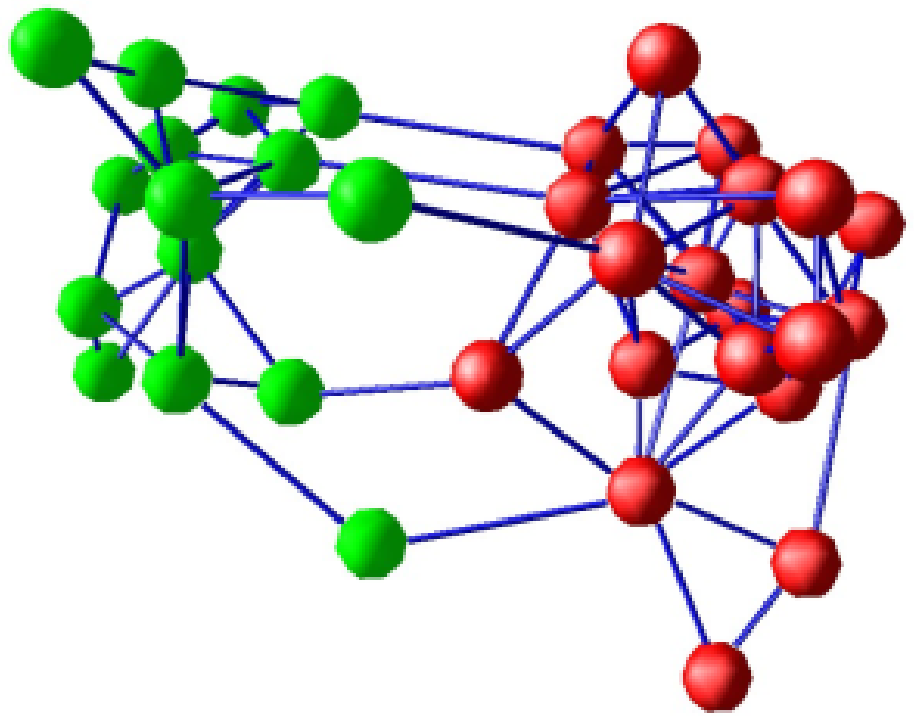,width=0.45\textwidth}
  \caption{
    (Color online) Network structure of a random network with 34 nodes, $\left<
    k \right> = 4.59$ and $p_{\mathrm{in}} / p_{\mathrm{out}} = 10$.
  }
  \label{fig:random_3}
\end{figure}

In Fig.\ \ref{fig:random_2} we show the modularity density for a random network
with 34 nodes and 78 undirected edges. Before distributing the edges, the nodes
were randomly assigned to 2 communities. The edge distribution was random with
the constraint that the probability $p_{\mathrm{in}}$ of connecting two nodes
within the same community was chosen to be 10 times larger than the probability
$p_{\mathrm{out}}$ (see Ref. \cite{PRE_69_026113} or Sec.
\ref{sec:general_networks} for a more detailed description of the
$p_{\mathrm{in}} / p_{\mathrm{out}}$ algorithm) of connecting two nodes which
are in different communities. The generated network is shown in Fig.\ 
\ref{fig:random_3} and it shows a clear community structure.

\begin{figure}
  \epsfig{file=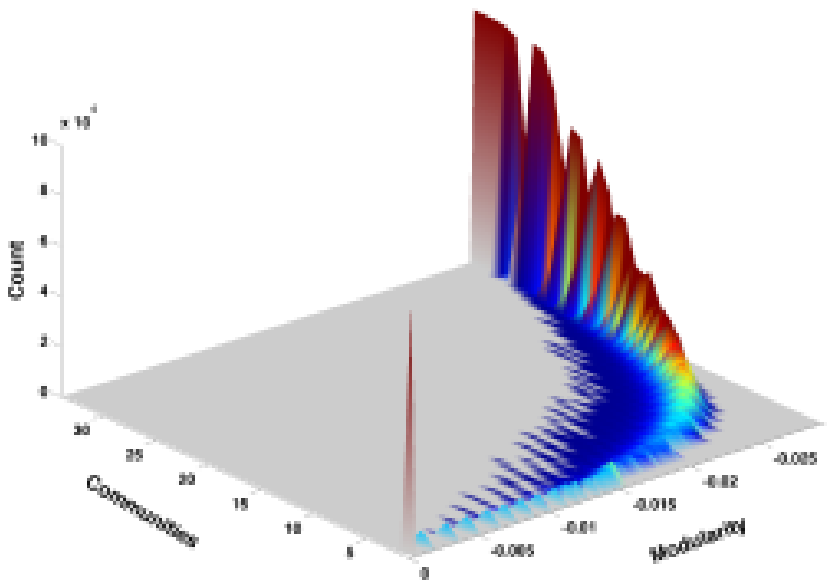,width=0.45\textwidth}
  \epsfig{file=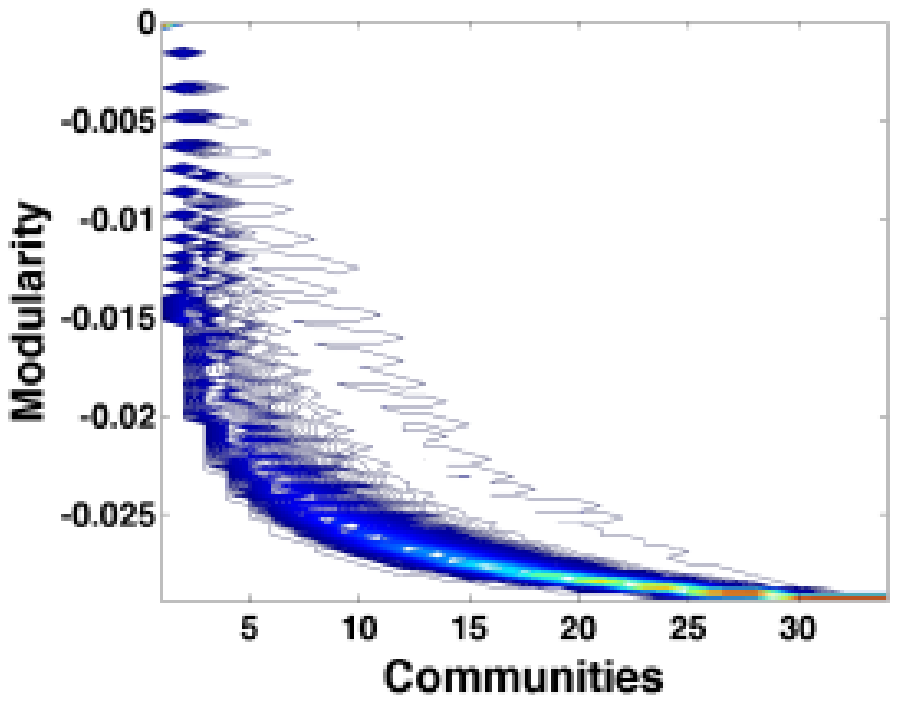,width=0.45\textwidth}
  \caption{
    (Color online) Modularity Density for a fully connected network with 34
    nodes.  On the x--axis is the number of communities, $C$.
  }
  \label{fig:full}
\end{figure}

\subsubsection{Fully connected network}

Fig.\ \ref{fig:full} shows our results for a fully connected network of 34
nodes.  The overall structure of the modularity density is markedly different
from the other cases and we attribute this difference to the much higher level
of connectedness of the nodes. In this case the modularity is always less than
zero and the maximum value is at $C = 1$, where $Q = 0$. The fact that the
modularity is negative is due to the fact that the number of inter--community
edges is always larger than the number of intra--community edges for this
network. The off--diagonal elements of the assortative mixing matrix $e$ are
therefore always large and contribute strongly to the negative term in the
expression for the modularity, eq.\ \ref{eq:modularity}. The average degree in
this fully connected network is $\left< k \right> = 33$ as compared to $\left< k
\right> = 4.59$ for the Zachary network.

\begin{figure}
  \epsfig{file=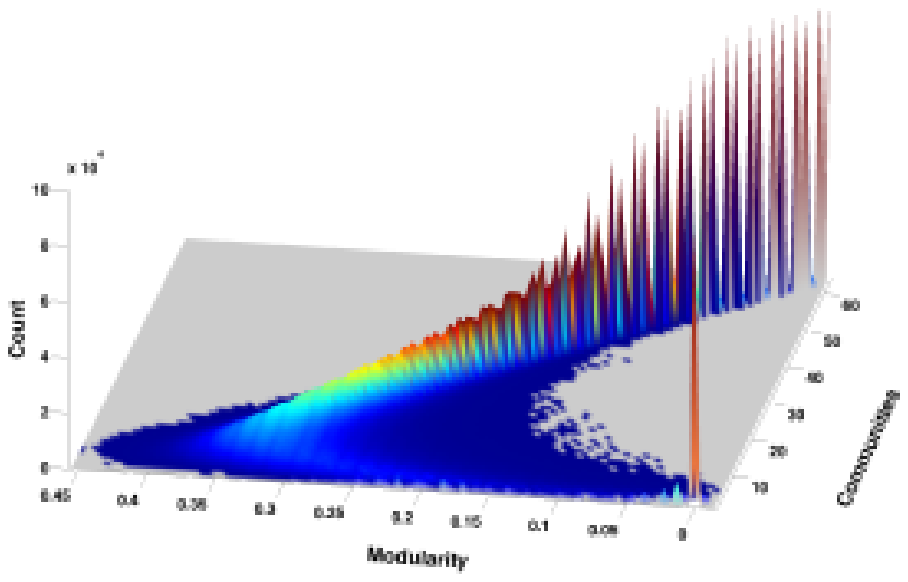,width=0.45\textwidth}
  \epsfig{file=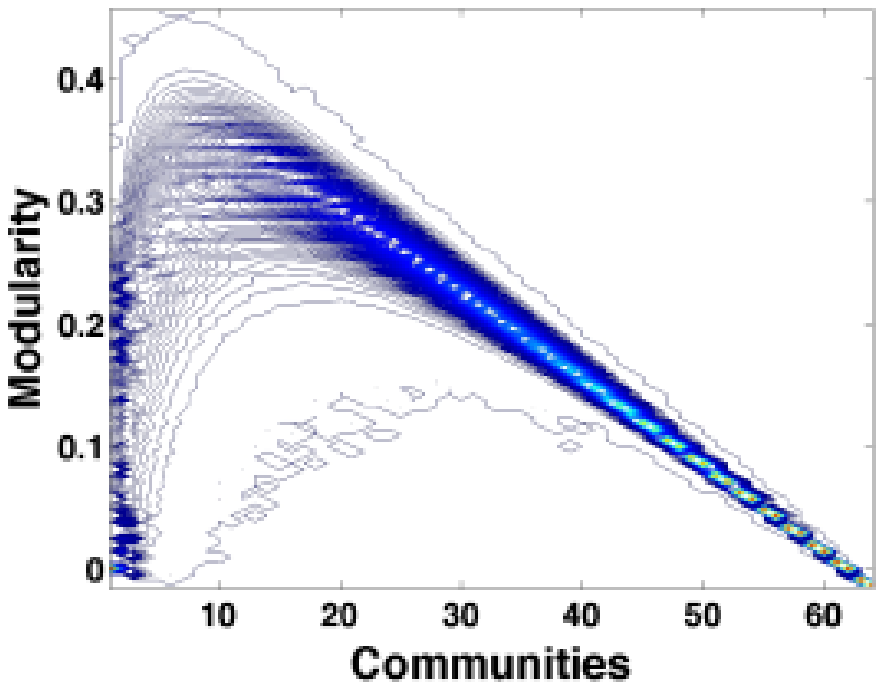,width=0.45\textwidth}
  \caption{
    (Color online) Modularity density for a regular network of $8 \times 8$
    nodes on a two dimensional square grid with periodic boundary conditions.
    On the x--axis is the number of communities, $C$.
  }
  \label{fig:symmetric}
\end{figure}

\subsubsection{Symmetric network}

In Fig.\ \ref{fig:symmetric} we show the modularity density of  a completely
symmetric regular network consisting of 64 nodes arranged in an $8 \times 8$
grid with 4 edges per node and periodic boundary conditions. Intuitively, this
network should not show any strong community structure. But as shown in the
figure, there is a high modularity region around $C \approx 8$ communities.

\subsection{Analysis}

\subsubsection{General networks}
\label{sec:general_networks}

The appearance of the modularity density is very similar for different networks
and in particular the shape of the most probable $Q(C)$ region, the ``ridge'' in
the density surface plots, shows remarkable similarity between different
networks.  In an effort to describe the general shape of the $Q(C)$ ridge we
will employ a simple model of the modularity as a function of the number of
communities. We start by observing that in order to maximize the modularity it
is desirable to minimize the number of off--diagonal elements and evenly
distribute the diagonal elements in the assortative mixing matrix $e$. In the
case where the communities are completely disconnected, the maximum modularity
is given by

\begin{equation}
  Q = 1 - \frac{1}{C},
\end{equation}

\noindent
where the first term is due to the trace of $e$ and the second is due to the
$\sum_{ij}(e^2)_{ij}$ term of eq.\ \ref{eq:modularity}.  This formula is the
upper theoretical limit of the modularity for any network but does not describe
the general shape of the modularity density very well. In order to make a
correction to this formula for networks that are connected we introduce the
function $\delta (C)$ which represents the average number of edges connecting a
community with the other communities. For simplicity we will only consider cases
where the communities have equal size and the edges between the communities are
distributed evenly. The $C \times C$ assortative mixing matrix will then look
like

\begin{equation}
  e_{C} =
    \frac{1}{M}\left(
      \begin{tabular}{llllllll}
        $\frac{M}{C}-\delta (C)$ & $\frac{\delta (C)}{C-1}$ & $\cdots$ \\
                                 &                          & \\
        $\frac{\delta (C)}{C-1}$ & $\frac{M}{C}-\delta (C)$ & \\
                                 &                          & \\
        $\vdots$                 &                          & $\ddots$
      \end{tabular}
    \right),
\end{equation}

\noindent
where $M$ is the total number of directed edges in the network. The modularity
in this case is given by

\begin{equation}
  Q = 1 - C\frac{\delta (C)}{M} - \frac{1}{C}.
\end{equation}

\noindent
In order to make a reasonable estimate of $\delta (C)$ we will revisit the
$p_{\mathrm{in}} / p_{\mathrm{out}}$--model already used in section
\ref{sec:random_networks}. The model, as introduced by \citet{PRE_69_026113}, is
used to generate networks with a predetermined community structure by randomly
choosing pairs of nodes and connecting them based on the two probabilities
$p_{\mathrm{in}}$ and $p_{\mathrm{out}}$. The ratio of the probabilities
determines the extent of community formation in the network. In addition, the
values of the probabilities are chosen such that the average degree per node,
$\left< k \right>$, can be controlled. Since we are interested in deriving a
simple closed form expression for $\delta(C)$ we approximate the probability of
finding a particular pair of nodes by assuming that the network is empty and
does not contain any edges. Our approximation will be good for sparsely
connected networks and few communities, but will become progressively worse for
$C \rightarrow N$ and $\left< k \right> \rightarrow N - 1$. In a network without
self--edges and $C$ communities we therefore require that

\begin{equation}
  \label{eq:k_condition}
  \left( \frac{N}{C} - 1 \right) p_{\mathrm{in}} +
    N \left( 1 - \frac{1}{C} \right) p_{\mathrm{out}} = \left< k \right>.
\end{equation}

\noindent
As a shorthand we introduce $\lambda = p_{\mathrm{in}} / p_{\mathrm{out}} \geq
1$, a freely tunable parameter. The first term in eq.\ (\ref{eq:k_condition})
corresponds to the average number of edges per node connecting two nodes within
the same community, $\left< k_{\mathrm{in}} \right>$. The second term is the
corresponding number of edges connecting two nodes in different communities,
$\left< k_{\mathrm{out}} \right>$. It is this second term that we need in order
to estimate the parameter $\delta (C)$. The parameter $\delta (C)$ is given by

\begin{equation}
  \delta (C) = \frac{N}{C} \left< k_{\mathrm{out}} \right>
    = \frac{N}{C} \left< k \right>
      \frac{ ( C - 1 ) }
           { ( 1 - C/N ) \lambda + ( C - 1 ) }.
  \label{eq:delta}
\end{equation}

\noindent
The fully connected like network can easily be derived now by setting $\lambda =
1$, and

\begin{equation}
  \label{eq:full_delta}
  \delta (C) = \left< k \right> \frac{N^{2}}{N-1} \frac{C-1}{C^{2}}.
\end{equation}

\begin{figure}
  \epsfig{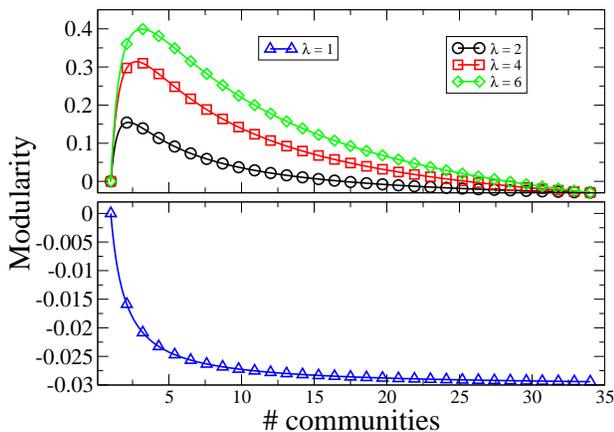}
  \caption{
    (Color online) Modularity plots for the $p_{\mathrm{in}} /
    p_{\mathrm{out}}$--model. The upper panel shows $\lambda = 2, 4, 6$ and the
    lower panel $\lambda = 1$. $N = 34$ and $\left< k \right> = 4.59$, which
    corresponds to the Zachary Karate Club.  On the x--axis is the number of
    communities, $C$.
  }
  \label{fig:QModel}
\end{figure}

\noindent
We chose $N = 34$ and $\left< k \right> = 4.59$ to model the Zachary Karate Club
and our results for a range of different values of $\lambda$ are shown in Fig.\
\ref{fig:QModel} in the upper and lower panels, respectively. As can be seen
from the figure, the model reproduces the shape of the most likely modularity,
the modularity density ridge, well for the networks shown in Figs.\
\ref{fig:Zachary-Surf} --
\ref{fig:random_2},
\ref{fig:full}, and
\ref{fig:symmetric}.
In particular the general feature of a peak in modularity at a small value of
$C$ is reproduced solely by the introduction of the $\delta(C)$ term that
describes the average number of edges that connect communities.  We observe that
any connected network can be approximated by an appropriate choice of
$\delta(C)$.  This suggests that any connected network will show a peak in
the modularity density for small $C \ll N$ just as our numerical results for real
networks indicate.  Note that the model is such that one needs to generate a new
network for each value of $C$ in order to keep $\lambda$ constant.
Consequently, the modularity plots in Fig.\ \ref{fig:QModel} are really to be
understood as a collection of modularity values for different networks given the
constraint that $\lambda$ is fixed. It is interesting to note also that the
fully connected network, shown in Fig.\ \ref{fig:full}, is qualitatively
reproduced by our model in the case of $\lambda = 1$ although $\left< k \right>$
is far from the fully connected value. In our model, $\left< k \right>$
represents only a scaling constant and does not alter the qualitative result. We
can not expect to get quantitative agreement since the model is only accounting
for the general behavior of the $C$--dependence of the modularity with fixed
$\lambda$. We note that it is in principle possible to write down the
$C$--dependence of $\lambda$ for a regular lattice but that has not been
performed in the current study. However, the limitations of the model do not
affect our general conclusion that the maximum in the modularity occurs for
relatively small values of $C$ for any connected network.




\section{Conclusions}
\label{section:Conclusions}

We have presented a matrix formalism to describe the modularity of a community
division in a network. We have described the modularity for some well studied
networks as well as some synthetic networks from a statistical point of view and
introduced the concept of modularity density.

In conclusion we found that the modularity density is quite similar for
different networks. Even random networks with no apparent community structure
exhibit a remarkably similar modularity density. This suggests that most of the
structure of the modularity density is independent of the network itself.  We
have introduced a simple model that describes the general shape of the
modularity density based on the $p_{\mathrm{in}} / p_{\mathrm{out}}$ concept of
\citet{PRE_69_026113} and concluded that any connected network must show a peak
in the modularity density at a small number of communities compared to the size
of the network.

The presence of a general shape indicates that it should be possible to develop
global optimization strategies which work well for most networks. The maximum
modularity curve of course depends on the particular network.


\section{Acknowledgments}

We would like to thank Travis Peery and Anders Niklasson for many helpful
discussions. We would also like to thank Mark Newman who graciously provided us
with his network data on the Zachary Karate Club and the Les Mis\'{e}rables
musical which we analyzed in this paper.  Special acknowledgments also to the
genuine and professional scientific atmosphere provided by the International Ten
Bar Caf\'{e}.

%
%
%
%
%

\bibliography{paper-20060825}

\begin{thebibliography}{19}
\expandafter\ifx\csname natexlab\endcsname\relax\def\natexlab#1{#1}\fi
\expandafter\ifx\csname bibnamefont\endcsname\relax
  \def\bibnamefont#1{#1}\fi
\expandafter\ifx\csname bibfnamefont\endcsname\relax
  \def\bibfnamefont#1{#1}\fi
\expandafter\ifx\csname citenamefont\endcsname\relax
  \def\citenamefont#1{#1}\fi
\expandafter\ifx\csname url\endcsname\relax
  \def\url#1{\texttt{#1}}\fi
\expandafter\ifx\csname urlprefix\endcsname\relax\def\urlprefix{URL }\fi
\providecommand{\bibinfo}[2]{#2}
\providecommand{\eprint}[2][]{\url{#2}}

\bibitem[{\citenamefont{Wilkinson and Huberman}(2004)}]{PNAS_101_5241}
\bibinfo{author}{\bibfnamefont{D.~M.} \bibnamefont{Wilkinson}}
  \bibnamefont{and} \bibinfo{author}{\bibfnamefont{B.~A.}
  \bibnamefont{Huberman}}, \bibinfo{journal}{Proc. Natl. Acad. Sci. USA}
  \textbf{\bibinfo{volume}{101}}, \bibinfo{pages}{5241} (\bibinfo{year}{2004}).

\bibitem[{\citenamefont{Massen and Doye}(2005)}]{PRE_71_046101}
\bibinfo{author}{\bibfnamefont{C.~P.} \bibnamefont{Massen}} \bibnamefont{and}
  \bibinfo{author}{\bibfnamefont{J.~P.~K.} \bibnamefont{Doye}},
  \bibinfo{journal}{Phys. Rev. E} \textbf{\bibinfo{volume}{71}},
  \bibinfo{pages}{046101} (\bibinfo{year}{2005}).

\bibitem[{\citenamefont{Meyers et~al.}(2005)\citenamefont{Meyers, Pourbohloul,
  Newman, Skowronski, and Brunham}}]{JTB_232_71}
\bibinfo{author}{\bibfnamefont{L.~A.} \bibnamefont{Meyers}},
  \bibinfo{author}{\bibfnamefont{B.}~\bibnamefont{Pourbohloul}},
  \bibinfo{author}{\bibfnamefont{M.~E.~J.} \bibnamefont{Newman}},
  \bibinfo{author}{\bibfnamefont{D.~M.} \bibnamefont{Skowronski}},
  \bibnamefont{and} \bibinfo{author}{\bibfnamefont{R.~C.}
  \bibnamefont{Brunham}}, \bibinfo{journal}{Journal of Theoretical Biology}
  \textbf{\bibinfo{volume}{232}}, \bibinfo{pages}{71} (\bibinfo{year}{2005}).

\bibitem[{\citenamefont{Leicht et~al.}(2005)\citenamefont{Leicht, Holme, and
  Newman}}]{physics_0510143}
\bibinfo{author}{\bibfnamefont{E.~A.} \bibnamefont{Leicht}},
  \bibinfo{author}{\bibfnamefont{P.}~\bibnamefont{Holme}}, \bibnamefont{and}
  \bibinfo{author}{\bibfnamefont{M.~E.~J.} \bibnamefont{Newman}},
  \bibinfo{journal}{physics} \textbf{\bibinfo{volume}{0510143}}
  (\bibinfo{year}{2005}).

\bibitem[{\citenamefont{Bagrow and Bollt}(2005)}]{PRE_72_046108}
\bibinfo{author}{\bibfnamefont{J.~P.} \bibnamefont{Bagrow}} \bibnamefont{and}
  \bibinfo{author}{\bibfnamefont{E.~M.} \bibnamefont{Bollt}},
  \bibinfo{journal}{PRE} \textbf{\bibinfo{volume}{72}}, \bibinfo{pages}{046108}
  (\bibinfo{year}{2005}).

\bibitem[{\citenamefont{Wu and Huberman}(2004)}]{EPJB_38_331}
\bibinfo{author}{\bibfnamefont{F.}~\bibnamefont{Wu}} \bibnamefont{and}
  \bibinfo{author}{\bibfnamefont{B.~A.} \bibnamefont{Huberman}},
  \bibinfo{journal}{The European Physical Journal B}
  \textbf{\bibinfo{volume}{38}}, \bibinfo{pages}{331} (\bibinfo{year}{2004}).

\bibitem[{\citenamefont{Reichardt and Bornholdt}(2004)}]{PRL_93_218701}
\bibinfo{author}{\bibfnamefont{J.}~\bibnamefont{Reichardt}} \bibnamefont{and}
  \bibinfo{author}{\bibfnamefont{S.}~\bibnamefont{Bornholdt}},
  \bibinfo{journal}{Phys. Rev. Lett.} \textbf{\bibinfo{volume}{93}},
  \bibinfo{pages}{218701} (\bibinfo{year}{2004}).

\bibitem[{\citenamefont{Ziv et~al.}(2005)\citenamefont{Ziv, Middendorf, and
  Wiggins}}]{PRE_71_046117}
\bibinfo{author}{\bibfnamefont{E.}~\bibnamefont{Ziv}},
  \bibinfo{author}{\bibfnamefont{M.}~\bibnamefont{Middendorf}},
  \bibnamefont{and} \bibinfo{author}{\bibfnamefont{C.~H.}
  \bibnamefont{Wiggins}}, \bibinfo{journal}{PRE} \textbf{\bibinfo{volume}{71}},
  \bibinfo{pages}{046117} (\bibinfo{year}{2005}).

\bibitem[{\citenamefont{Newman}(2002)}]{PRL_89_208701}
\bibinfo{author}{\bibfnamefont{M.~E.~J.} \bibnamefont{Newman}},
  \bibinfo{journal}{Phys. Rev. Lett.} \textbf{\bibinfo{volume}{89}},
  \bibinfo{pages}{208701} (\bibinfo{year}{2002}).

\bibitem[{\citenamefont{Newman}(2003)}]{PRE_67_026126}
\bibinfo{author}{\bibfnamefont{M.~E.~J.} \bibnamefont{Newman}},
  \bibinfo{journal}{Phys. Rev. E} \textbf{\bibinfo{volume}{67}},
  \bibinfo{pages}{0261126} (\bibinfo{year}{2003}).

\bibitem[{\citenamefont{Newman and Girvan}(2004)}]{PRE_69_026113}
\bibinfo{author}{\bibfnamefont{M.~E.~J.} \bibnamefont{Newman}}
  \bibnamefont{and} \bibinfo{author}{\bibfnamefont{M.}~\bibnamefont{Girvan}},
  \bibinfo{journal}{Phys. Rev. E} \textbf{\bibinfo{volume}{69}},
  \bibinfo{pages}{026113} (\bibinfo{year}{2004}).

\bibitem[{\citenamefont{Girvan and Newman}(2002)}]{PNAS_99_7821}
\bibinfo{author}{\bibfnamefont{M.}~\bibnamefont{Girvan}} \bibnamefont{and}
  \bibinfo{author}{\bibfnamefont{M.~E.~J.} \bibnamefont{Newman}},
  \bibinfo{journal}{Proc. Natl. Acad. Sci. USA} \textbf{\bibinfo{volume}{99}},
  \bibinfo{pages}{7821} (\bibinfo{year}{2002}).

\bibitem[{\citenamefont{Clauset et~al.}(2004)\citenamefont{Clauset, Newman, and
  Moore}}]{PRE_70_066111}
\bibinfo{author}{\bibfnamefont{A.}~\bibnamefont{Clauset}},
  \bibinfo{author}{\bibfnamefont{M.~E.~J.} \bibnamefont{Newman}},
  \bibnamefont{and} \bibinfo{author}{\bibfnamefont{C.}~\bibnamefont{Moore}},
  \bibinfo{journal}{Phys. Rev. E} \textbf{\bibinfo{volume}{70}},
  \bibinfo{pages}{066111} (\bibinfo{year}{2004}).

\bibitem[{\citenamefont{Newman}(2004)}]{PRE_69_066133}
\bibinfo{author}{\bibfnamefont{M.~E.~J.} \bibnamefont{Newman}},
  \bibinfo{journal}{Phys. Rev. E} \textbf{\bibinfo{volume}{69}},
  \bibinfo{pages}{066133} (\bibinfo{year}{2004}).

\bibitem[{\citenamefont{Medus et~al.}(2005)\citenamefont{Medus, Acu\~{n}a, and
  Dorso}}]{Physica_A_358_593}
\bibinfo{author}{\bibfnamefont{A.}~\bibnamefont{Medus}},
  \bibinfo{author}{\bibfnamefont{G.}~\bibnamefont{Acu\~{n}a}},
  \bibnamefont{and} \bibinfo{author}{\bibfnamefont{C.~O.} \bibnamefont{Dorso}},
  \bibinfo{journal}{Physica A} \textbf{\bibinfo{volume}{358}},
  \bibinfo{pages}{593} (\bibinfo{year}{2005}).

\bibitem[{\citenamefont{Guimera and Amaral}(2005)}]{Nature_433_895}
\bibinfo{author}{\bibfnamefont{R.}~\bibnamefont{Guimera}} \bibnamefont{and}
  \bibinfo{author}{\bibfnamefont{L.}~\bibnamefont{Amaral}},
  \bibinfo{journal}{Nature} \textbf{\bibinfo{volume}{433}},
  \bibinfo{pages}{895} (\bibinfo{year}{2005}).

\bibitem[{\citenamefont{Duch and Arenas}(2005)}]{Duch_05}
\bibinfo{author}{\bibfnamefont{J.}~\bibnamefont{Duch}} \bibnamefont{and}
  \bibinfo{author}{\bibfnamefont{A.}~\bibnamefont{Arenas}},
  \bibinfo{journal}{Phys. Rev. E} \textbf{\bibinfo{volume}{72}},
  \bibinfo{pages}{027104} (\bibinfo{year}{2005}).

\bibitem[{\citenamefont{Zachary}(1977)}]{J_Anthropol_Res_33_452}
\bibinfo{author}{\bibfnamefont{W.~W.} \bibnamefont{Zachary}},
  \bibinfo{journal}{J. Anthropol. Res.} \textbf{\bibinfo{volume}{33}},
  \bibinfo{pages}{452} (\bibinfo{year}{1977}).

\bibitem[{Les()}]{LesMis}
\bibinfo{note}{\protect{www.lesmis.com}}.

\end{thebibliography}
\bibliographystyle{apsrev}

\end{document}